\documentclass[amsmath,amssymb,aps,prb,lengthcheck,superscriptaddress,twocolumn]{revtex4}
\usepackage[T1]{fontenc}
\usepackage[latin1]{inputenc}
\usepackage{amsmath}
\usepackage{graphicx}
\usepackage{epsfig}
\usepackage{bm}
\usepackage{dsfont}
\usepackage{color}
\usepackage{braket}
\usepackage{bbm}
\usepackage{physics}

\newcommand{\bea}{\begin{eqnarray}}
\newcommand{\ea}{\end{eqnarray}}
\newcommand{\bc}{\begin{center}}
\newcommand{\ec}{\end{center}}

\begin{document}

\author{Damian Wozniak}
 \affiliation{Institut f\"ur Physik, Universit\"at Greifswald, 17487 Greifswald, Germany}
 \affiliation{Department of Physics, Royal Holloway, University of London,
 Egham, Surrey TW20 0EX, United Kingdom}
\title{Chaos onset in large rings of Bose-Einstein condensates}
\author{Johann Kroha}
\affiliation{Fachbereich Physik, Universit\"at Bonn, D-53115 Bonn, Germany}
\author{Anna Posazhennikova}
\affiliation{Institut f\"ur Physik, Universit\"at Greifswald, 17487 Greifswald, Germany}

\date{\today}
\begin{abstract}

We consider large rings of weakly-coupled Bose-Einstein condensates, analyzing their transition to chaotic dynamics and loss of coherence. Initially, a ring is considered to be in an eigenstate, i.e. in a commensurate configuration with equal site fillings and equal phase differences between neighboring sites.  Such a ring should exhibit a circulating current whose value will depend on the initial, non-zero phase difference. The appearance of such currents is a signature of an established coherence along the ring. If phase difference falls between $\pi/2$ and $3\pi/2$ and interparticle interaction in condensates exceeds a critical interaction value $u_c$, the coherence is supposed to be quickly destroyed because the system enters a chaotic regime due to inherent instabilities. This is, however, only a part of the story. It turns out that chaotic dynamics and resulting averaging of circular current to zero is generally offset by a critical time-scale $t_c$, which is almost two orders of magnitude larger than the one expected from the linear stability analysis.
 We study the critical time-scale in detail in a broad parameter range. 

\end{abstract}



\maketitle

\section{Introduction}

Ring-coupled Bose-Einstein condensates (BECs) with an initially finite phase difference between neighboring sites constitute a particularly interesting system. They allow for circulating currents which results in a controlled formation of topological defects such as vortices. Possible applications of such systems range from interferometry \cite{Kasevich1997,Anderson1998} to quantum computation, atomtronics and SQUIDS \cite{Hallwood2010,Amico2014,Amico2015,Arwas2016,Ryu2013}.   

A while ago, a ring of three condensates was studied in the whole range of initial phase differences between neighboring sites \cite{Tsubota2000,Tsubota2002} with the goal to find the probability of vortex generation via the Kibble-Zurek mechanism in nonuniform, domain-structured superfluids. Experimentally the idea was tested by three ${}^{87}$Rb condensates merging, which indeed led to the formation of vortices, whose number strongly depended on the merging velocity \cite{Scherer2007}.

Although in a ring of three coupled condensates, circular current can be nonzero when all three phase differences differ from each other, the maximum value of the circular current is reached only for the commensurate case, i.e. when all the three phase differences are the same \cite{Tsubota2000}. This is because only a commensurate case corresponds to an eigenstate of the system.  Importantly, this circular current depends on the system parameters, in particular on the interaction between condensed particles. Linear stability analysis provides a critical  interaction value $u_c$, above which some of the eigenmodes become unstable and chaotic dynamics sets in for $u>u_c$. The time-averaged circular current gradually tends to zero as the interaction increases apart from the remaining sharp peaks associated with the eigenmodes \cite{Tsubota2000}.  It is therefore not clear why nonzero circular currents are still present for $u\gg u_c$ in the numerical results of Ref. \cite{Tsubota2000}.

Further studies on condensate rings (with number of sites $N_s \geq 3$) investigated different theoretical aspects including dynamical and thermodynamical stability \cite{Paraoanu2003,dePassos2009}, chaos and ergodicity \cite{Arwas2017,Arwas2019}, symmetry analysis and effects of quantum many-body dynamics \cite{Nemoto2000} and quantum quenches \cite{Zurek2011}. However, in those works, the values of the circular currents were not investigated in detail in the chaotic regime, and time scales associated with the currents were not discussed. 

From an experimental point of view, it is rather challenging to realize circulating currents in large rings and for large winding numbers due to their quick decay to flows with lesser winding numbers (see for instance experiments on ${}^{87}$Rb annular condensates in Ref. \cite{Hadzibabic2012}). Recently,  a substantial progress in the creation of stable superflows has been achieved in six- and seven-site rings of polaritonic condensates confined in microcavities \cite{Cookson2021}. Particularly interesting is that persistent circular currents with large winding numbers have been observed for nominally unstable initial configurations for $N_s=7$ and winding numbers $k=2$ and $k=3$.\cite{Cookson2021}

Motivated by these developments and open questions, we investigate the effect of chaos on circular currents in detail, i.e. dependence on interaction, initial conditions and system size. We show that, although chaos obstructs circular flow, it does not set in immediately even if the system is tuned to an unstable eigenmode. We identify a critical time scale associated with this type of dynamics and demonstrate how the time scale depends on the systems various parameters. We show that the time scale is much larger than the one expected from the linear stability analysis, which explains numerical results of Ref. \cite{Tsubota2000}, and can provide an insight into experimental results of Ref. \cite{Cookson2021}. 

The paper is organized as follows: In section \ref{Model} we formulate the model, derive equations of motion and the expression for circular current.  In section \ref{noninteracting} we analyze the circular current in the non-interacting system for a special case of symmetric initial conditions and show the current becomes a simple sine wave in the limit of large $N_s$.  
In section \ref{inter} after a brief discussion of unstable modes and characteristic interaction, we analyze time-dependent circular current at unstable modes and identify a particular time scale $t_c$ associated with system transition to a chaotic regime. We show that the slide to the chaotic regime occurs exponentially and study the time-averaged circular current showing how the current at unstable modes gradually disappear upon increasing interaction. We conclude in section \ref{last}. 

\section{Model and equations of motion}

\label{Model}

We consider a system of $N$ condensed bosons trapped in a one-dimensional periodic potential consisting of $N_s$ wells with periodic boundary conditions. The average filling factor in the system
\begin{equation}
    \rho=\frac{N}{N_s}
\end{equation}
 is assumed to be macroscopic so that the semiclassical Gross-Pitaevskii approximation is applicable for the system description (for example, in experiments on long arrays of ${}^{87}$Rb condensates $\rho\approx 1000$ according to Ref.  \cite{Cataliotti2001}). The local condensates are considered to be weakly linked in order for Josephson current to be induced between the sites \cite{Smerzi1997}. Note that in all calculations we keep the number of particles per site, $\rho$, constant when changing the system size $N_s$. The general Gross-Pitaevskii equation reads 
\begin{equation}
\mathrm{i}\hbar \frac{\partial}{\partial t} \Psi({\bf r},t)=\left( -\frac{\hbar^2}{2m}\nabla^2+V_{ext}({\bf r})+g | \Psi({\bf r},t)|^2\right)\Psi({\bf r},t),
\label{GPeq}
\end{equation}
where $\Psi({\bf r},t)$ is the mean-field averaged bosonic field operator $ \langle \hat \Psi({\bf r},t)  \rangle $, $V_{ext}({\bf r})$ is the multi-well external potential, $g$ is a repulsive contact interaction constant. By expanding the semi-classical wave function in terms of a set of localized basis functions (Wannier functions $\Phi_i({\bf r}$),
\begin{equation}
\Psi({\bf r},t)=\sum_{i=1}^{N_s}\phi_i({\bf r})\psi_i(t),
\label{multiwell}
\end{equation}
and integrating out the spatial degrees of freedom, 
we obtain the standard discrete nonlinear Schr\"odinger equations (DNLSEs) for $\psi_i$-s \cite{Trombettoni2001}
\begin{eqnarray}
\mathrm{i}\hbar \frac{\partial}{\partial t}\psi_i(t)=[E_i+U_i|\psi_i (t)|^2]\psi_i(t)-K_{i,i-1}\psi_{i-1}(t) \nonumber \\ -K_{i,i+1}\psi_{i+1}(t), \quad i=1,2,...,N_s.
\label{GPsystem}
\end{eqnarray}
The  periodic boundary conditions imply ($N_s+1\rightarrow1$). DNLSEs adequately capture the dynamics of multiple coupled condensates, as was verified in the experimental work \cite{Cataliotti2001}. 
\begin{figure}[!bt]
\begin{center}
\includegraphics[width=0.30\textwidth]{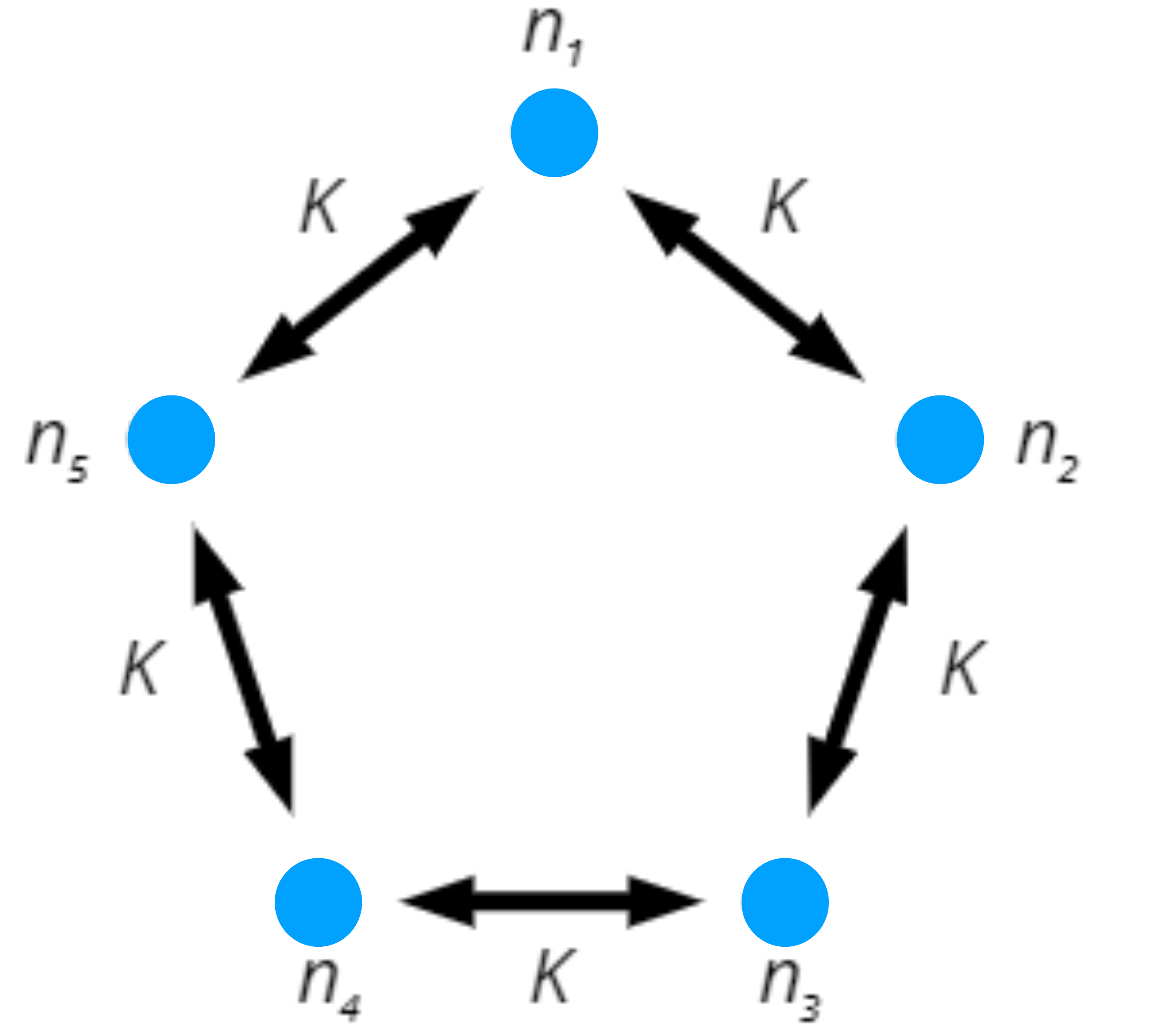} 
\end{center}
\caption{  Schematic setup of the ring system for five wells ($N_s=5$). Blue circles represent condensates wave-functions, $K$ is the Josephson coupling constant \eqref{param}, $n_i$-s are condensate populations according to  \eqref{population}. }
\end{figure} 
The model parameters in Eq.~(\ref{GPsystem}), the  zero-point energies $E_i$, on-site interaction $U_i$ and Josephson couplings between neighbouring wells $K_{i,i\pm1}$ are given by \cite{Smerzi1997}
\begin{eqnarray}
E_i&=& \int d{\bf r} \left[ \frac{\hbar^2}{2m} |\nabla \phi_i({\bf r})|^2+|\phi_i({\bf r})|^2 V_{ext}({\bf r}) \right], \nonumber \\
U_i&=&g\int d{\bf r} |\phi_i({\bf r})|^4, \nonumber \\
K_{i,i\pm 1}&=&-\int d{\bf r}\left[ \frac{\hbar^2}{2m}\nabla \phi_i({\bf r})\nabla \phi_{i\pm1}({\bf r}) \right. \nonumber \\ &+&\left. \phi_i({\bf r})V_{ext}({\bf r})\phi_{i\pm1}({\bf r}) \right].
\label{param}
\end{eqnarray}
We make use of the ansatz 
\begin{equation}
\psi_i(t)=\sqrt{n_i}(t)e^{\mathrm{i}\theta_i(t)}
\label{population}
\end{equation}
where we introduced the site populations $n_i$  normalized by the filling factor $\rho$
\begin{equation}
    n_i(t)=\frac{N_i(t)}{\rho},
\end{equation}
where $N_i(t)$ is the number of particles in the condensate on site $i$ at time $t$. Thus, an initially homogeneous distribution of atoms means $n_i(0)=1$ for all $i$.
With that Eqs. \eqref{GPsystem} can be rewritten as a set of differential equations for $n_i$ and phase differences $\theta_{i,i+1}=\theta_{i+1}-\theta_i$ as follows
\begin{eqnarray}
\dot n_i&=&-2\sqrt{n_i}(\sqrt{n_{i+1}}\sin\theta_{i,i+1}-\sqrt{n_{i-1}}\sin\theta_{i-1,i}), \nonumber \\
\dot \theta_{i,i+1}&=&u(n_{i}-n_{i+1})+\left(\sqrt{\frac{n_{i}}{n_{i+1}}}- \sqrt{\frac{n_{i+1}}{n_{i}}}\right)\cos\theta_{i,i+1} \nonumber \\
&+&\sqrt{\frac{n_{i-1}}{n_i}}\cos\theta_{i-1,i}-\sqrt{\frac{n_{i+2}}{n_{i+1}}}\cos\theta_{i+1,i+2}. 
\label{eqs_fin}
\end{eqnarray}
In deriving these equations we assumed the following simplifications: $E_i=0$, $U_i\equiv U$ and $K_{i,i+1}\equiv K$ for all values of $i$. We also expressed the time argument $t$ in units of $\hbar/K$, and introduced the dimensionless interaction parameter
\begin{equation}
u=\frac{U \rho}{ K}. 
\label{int_dim}
\end{equation}
These equations conserve $\sum_i n_i=N_s$ and the total energy
\begin{equation}
    E=\frac{\rho K}{\hbar} \left( \frac{u}{2}\sum_i n_i^2-2 \sum_{i} \sqrt{n_in_{i+1}}\cos \theta_{i,i+1}\right).
    \label{energy}
\end{equation}
We define the circular current as the average current in a clockwise loop around the ring of condensates
\begin{equation}
I=\frac{1}{N_s} \sum_{i=1}^{N_s}I_{i,i+1},
\label{curr_ns}
\end{equation}
where $I_{i,i+1}$ is the particle current from site $i$ to $i+1$ defined as 
\begin{equation}
I_{i,i+1}=2\operatorname{Im} (\psi_i^*(t)\psi_{i+1}(t))=2\sqrt{n_in_{i+1}}\sin(\theta_{i,i+1}).
\label{curr_psi}
\end{equation}
Note that the current defined in this way has units $\rho K/\hbar$. In the next section, we analyze the current analytically for $u=0$ and a special case of initial conditions.

\section{Circular current in the non-interacting case}
\label{noninteracting}

The non-interacting case with $U_i=0$ and coupling constants $K_{i,i+1}\equiv K$ can be solved exactly for any number of sites $N_s$, as shown in Appendix A. However, one does not need these solutions in order to calculate circular current, as it can be straightforwardly derived from the current conservation condition (one can show, for example, that $\partial _t I=0$ with the help of Eqs.\eqref{GPsystem}). We thus get for the circular current 
\begin{equation}
I=I_0=\frac{2}{N_s}  \operatorname{Im}\sum_{i=1}^{N_s}\psi_i^*(0)\psi_{i+1}(0). \label{curr_nonint}
\end{equation} 
We see that the current is constant and the value of this constant depends on initial $n_i(0)$-s and $\theta_{i,i+1}(0)$-s. For the homogeneous condensate distribution $n_i(0)=1$, the current only depends on the initial phase differences. 

\begin{figure}[!bt]
\begin{center}
\includegraphics[width=0.47\textwidth]{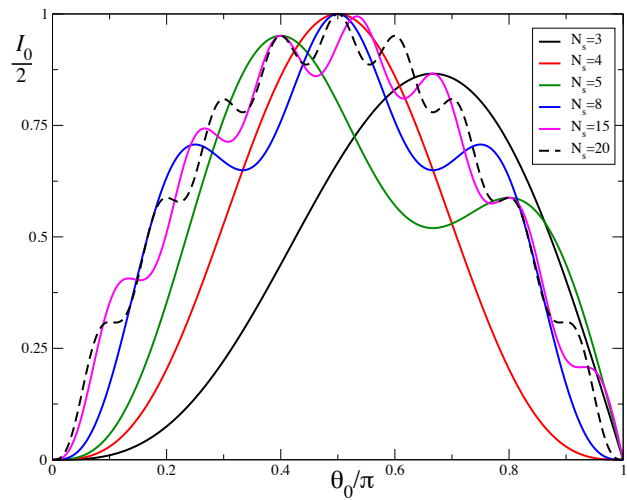} 
\end{center}
\caption{  Average  current of the noninteracting system $I_0$ \eqref{current_zero} divided by its maximum value (=2)  versus initial phase difference $\theta_0$ for initial conditions \eqref{initial}, plotted for condensate rings of different sizes  ( $N_s=3;4;5;8;15$ and $20$). }
\label{zero_current}
\end{figure} 

We chose the following initial conditions for the phase differences
\begin{equation}
\theta_{i,i+1} = \left\{ \begin{array}{ccl}
\theta_{0} & \mbox{for} & i=1,...,N_s-1, \\
-(N_s-1)\theta_{0} & \mbox{for} & i=N_s,
\end{array}\right.
\label{initial_pahse_diff}
\end{equation}
so that
\begin{equation}
\psi_j(0)=e^{\mathrm{i}(j-1)\theta_0},  \quad j=1,2,..,N_s,
\label{initial}
\end{equation}

We now get for the current in Eq. \eqref{curr_nonint} the simple expression
\begin{equation}
I_0(\theta_0)=\frac{2}{N_s}\left\{ (N_s-1)\sin\theta_{0}-\sin[(N_s-1)\theta_{0}] \right\}.
\label{current_zero}
\end{equation}

Since the current is an odd function of the initial phase difference $I_{av}^0(\theta_0)=-I_{av}^0(-\theta_0)$ and is $2\pi$-periodic, it is sufficient to consider $\theta_0 \in [0,\pi]$. In Fig. \ref{zero_current} we plotted $I_0$ normalized by its maximum value $(I_0)_{max}=2$ versus $\theta_0$ for various ring sizes. 
We see that the current has local maxima at discrete values of $\theta_0$ given by 
 \begin{equation}
 \theta_0^{k}=\frac{2\pi k}{N_s},
 \label{loc_max}
\end{equation}
where the integer $k=1,...,N_s$ has the meaning of the winding number. This result is not surprising and could be alternatively found from the extrema of the total energy of the noninteracting system
\eqref{energy}
\begin{equation}
E_0=-2\frac{\rho K}{\hbar} \left[ (N_s-1)\cos\theta_0+\cos[(N_s-1)\theta_0 ] \right].
\label{energy_nonint}
\end{equation}
This is because $\theta_0^k$-s play the role of quantized components of the effective quasi-momentum in Fourier space, and the group velocity and as a consequence the circular current is proportional to $\partial E/\partial \theta_0^k$. Naturally, the energy values at \eqref{loc_max} coincide (up to our normalization factor) with the eigenvalues $\lambda_k$ of noninteracting Hamiltonian calculated in Appendix A and read
\begin{equation}
    E_k=-2 \frac{N K}{\hbar} \cos \frac{2\pi k}{N_s}.
\end{equation}

Note that the discrete modes  \eqref{loc_max}  correspond to the commensurate case with homogeneous initial conditions. Values of $\theta_0$ in between the discrete modes belong to a special case of incommensurate configuration when all but one (the phase difference between first and the last sites) initial phase differences are the same as shown in Eq.\eqref{initial_pahse_diff}.  

At the discrete modes \eqref{loc_max} the current \eqref{current_zero} is simply a sine-function
\begin{equation}
I_0(\theta_0^{k})=2 \sin\left( \theta_0^{k}  \right). 
\label{current_zero_simplified}
\end{equation}

\section{Circular current for ring-coupled, interacting condensates}

\label{inter}

\subsection{Chaotic behavior and the critical interaction}

\label{u_crit}

Once the interaction is switched on, the system dynamics become more complicated, and analytical solutions are no longer possible. It is important that, contrary to the non-interacting case, the dynamics become chaotic for a specific parameter range. 
To identify the parameter range, we start from the linear stability analysis of the coupled real Eqs.  \eqref{eqs_fin}. The stability is decided by the eigenvalues of the corresponding Jacobian matrix (see Appendix B for details).  These $2N_s$ eigenvalues can be derived analytically due to the Jacobian matrix's special, block-wise circulant structure. The eigenvalues are
\begin{equation} \label{eq1}
\begin{split}
\lambda_{j}(k)  & = 2 \mathrm{i} \left\{  \sin \theta_0^{k}  \sin{ \left(\frac{2 \pi j } {N_s} \right) }  
\pm  \right.   \\
 & \left. \sin{ \left(\frac{ \pi j } {N_s} \right) } \sqrt{ 2 \cos\theta_0^{k} \left[2\cos\theta_0^{k} \sin^2\left( \frac{\pi j}{N_s} \right) + u  \right]    } \right\}, 
\end{split}
\end{equation}
where $j=1,2,..,N_s$.  The eigenvalues depend on the eigenmodes $\theta_0^{k}$, and the dimensionless interaction $u$ from Eq.\eqref{int_dim}.
Note that $\epsilon_j(k)=-\mathrm{i}\lambda_j(k)$ correspond to the discrete Bogoliubov excitation spectrum of the system, discussed previously in the context of condensate arrays \cite{Smerzi2002}, and the Bogoliubov-de Gennes description of a circular array of Bose-Einstein condensates \cite{Paraoanu2003}.
The eigenvalues $\lambda_j$ are zero or purely imaginary, and both Eqs.~\eqref{eqs_fin} are stable unless the expression under the square root turns negative. Since we consider only repulsive interactions, $u>0$, the condition for at least one eigenvalue to acquire a real part is 
\begin{eqnarray}
\cos\theta_0^{k}<0, \label{cond_ini} \\
2\cos\theta_0^{k} \sin^2\left( \frac{\pi j}{N_s} \right) + u>0. 
\end{eqnarray}
The analysis of these inequalities provides the expression for the upper bound of interaction below which the stationary points of the nonlinear equations are still stable. The first eigenvalue acquiring a real part turns the system unstable, which first occurs for $j=1$, giving us the critical interaction
 \begin{equation}
u_c=-2\cos\theta_0^k \, \sin^2\left( \frac{\pi }{N_s} \right).
\label{uc}
\end{equation}
Note that $u_c>0$ because of the condition \eqref{cond_ini}.
We see that $u_c$ depends on the initial conditions through the modes $\theta_0^k$, with the mode selection criterion Eq. \eqref{cond_ini}. Accordingly, the range of unstable modes is defined by the interval
 \begin{equation}
 \frac{\pi}{2}<\theta_0^k<\frac{3 \pi}{2} \ \ \text{or} \ \ N_s<4 k<3N_s.  
\label{pos_modes}
\end{equation}
We will refer to such modes as "unstable discrete modes", keeping in mind that they become unstable for $u>u_c$. Conditions similar to Eqs. \eqref{uc}, \eqref{pos_modes} were derived also in Ref. [\onlinecite{Paraoanu2003}] from the Bogoliubov-de Gennes equations.
Conversely, modes in the complementary range, $$\theta_0^k\in [-\pi/2, \pi/2]$$ are stable and will be called "stable discrete modes".

\begin{figure}[!b]
\begin{center}
\includegraphics[width=0.47\textwidth]{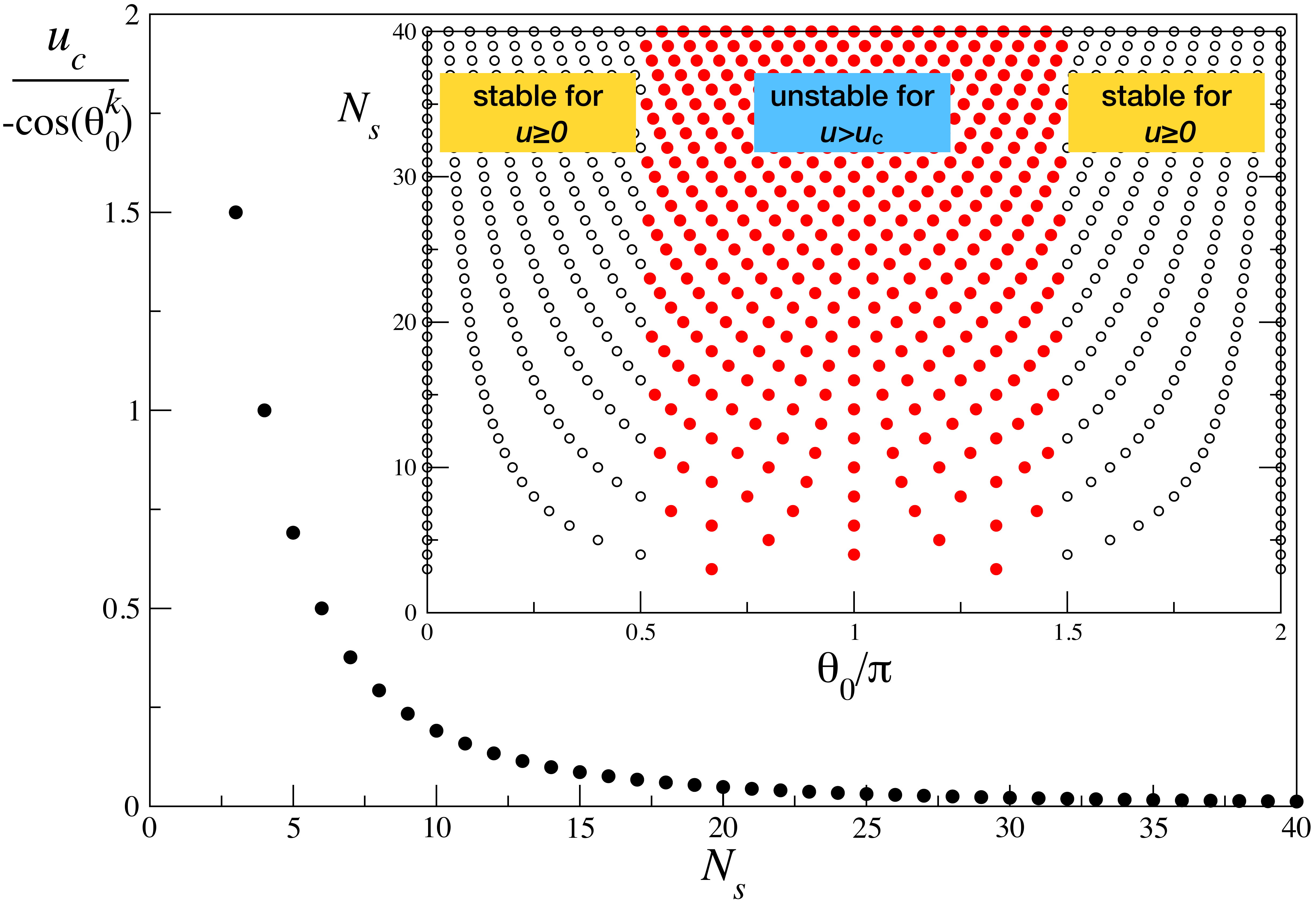} 
\end{center}
\caption{Dependence of characteristic interaction $u_c$ (divided by  $\cos(\theta_0^k)$) on the number of sites $N_s$, ranging between 3 and 40. Inset: stability diagram showing  discrete modes, i.e. values of initial phase difference $\theta_0$ from \eqref{loc_max} versus $N_s$. The modes denoted by open circles represent modes that are stable independently of the interaction $u$-value. Discrete modes denoted by solid circles showcase the modes which are stable for $0<u\le u_c$ but become unstable for $u>u_c$.}
\label{uc_modes}
\end{figure}

It follows from Eqs.\eqref{uc} and \eqref{pos_modes} that the maximum possible value of the critical interaction $u_c=1$, and it is reached for the $\pi-$mode of a four-site ring. Since $u_c$ is just proportional to the corresponding cosine of the corresponding discrete mode, one can plot a universal, mode-independent graph for normalized characteristic interaction $u_c/(-\cos\theta_0^k)$, which we display in Fig. \ref{uc_modes}. One can see that $u_c$ tends to zero relatively fast with the increasing number of sites $N_s$. The inset of the graph shows the stability diagram, where modes denoted by open circles represent stable solutions independent of the value of $u$ (given that only non-negative interactions are considered), whereas solid circles represent modes that become unstable for $u>u_c$. 

The linear stability analysis introduces the instability exponent $\alpha_0$, given by the real part of the Jacobian eigenvalue $\lambda_1$,
\begin{equation}
    \alpha_0\equiv  Re (\lambda_1)=2 \sin{ \left(\frac{ \pi } {N_s} \right) } \sqrt{ 2 |\cos\theta_0^{k}| \left[u-u_c  \right]    }.
    \label{alpha_analytical}
\end{equation}
This $\alpha_0$ is the rate by which the circular current $I(t)$ of a given mode $\theta_0^k$ is expected to deviate exponentially in time from its stationary value $I_0(\theta_0^k)$. To be more specific, it should be $2\alpha_0$, since the current is proportional to the product of two condensate wave functions. The equation \eqref{alpha_analytical} hence establishes  the deviation time scale 
$\sim 1/\alpha_0$ which would diverge at the stability boundaries, i.e., for 
$\theta_0 \to \pi/2$ or $3\pi/2$, or $u \to u_c$. We will show in the following section that, interestingly, 
chaotic behavior sets in abruptly at a much larger, critical time $t_c\gg 1/\alpha_0$, due to the nonlinear character of the system, not captured by the 
linear stability analysis.

\subsection{Temporal onset of chaotic behavior }

In this section, we fully characterize the decay of the unstable, discrete modes by studying the circulating current numerically beyond linear stability analysis. Initially, the circular current has a nonzero value equal to its noninteracting value $I_0$, derived in section \ref{noninteracting}. 

\begin{figure}[t!]
\begin{center}
\includegraphics[width=0.48\textwidth]{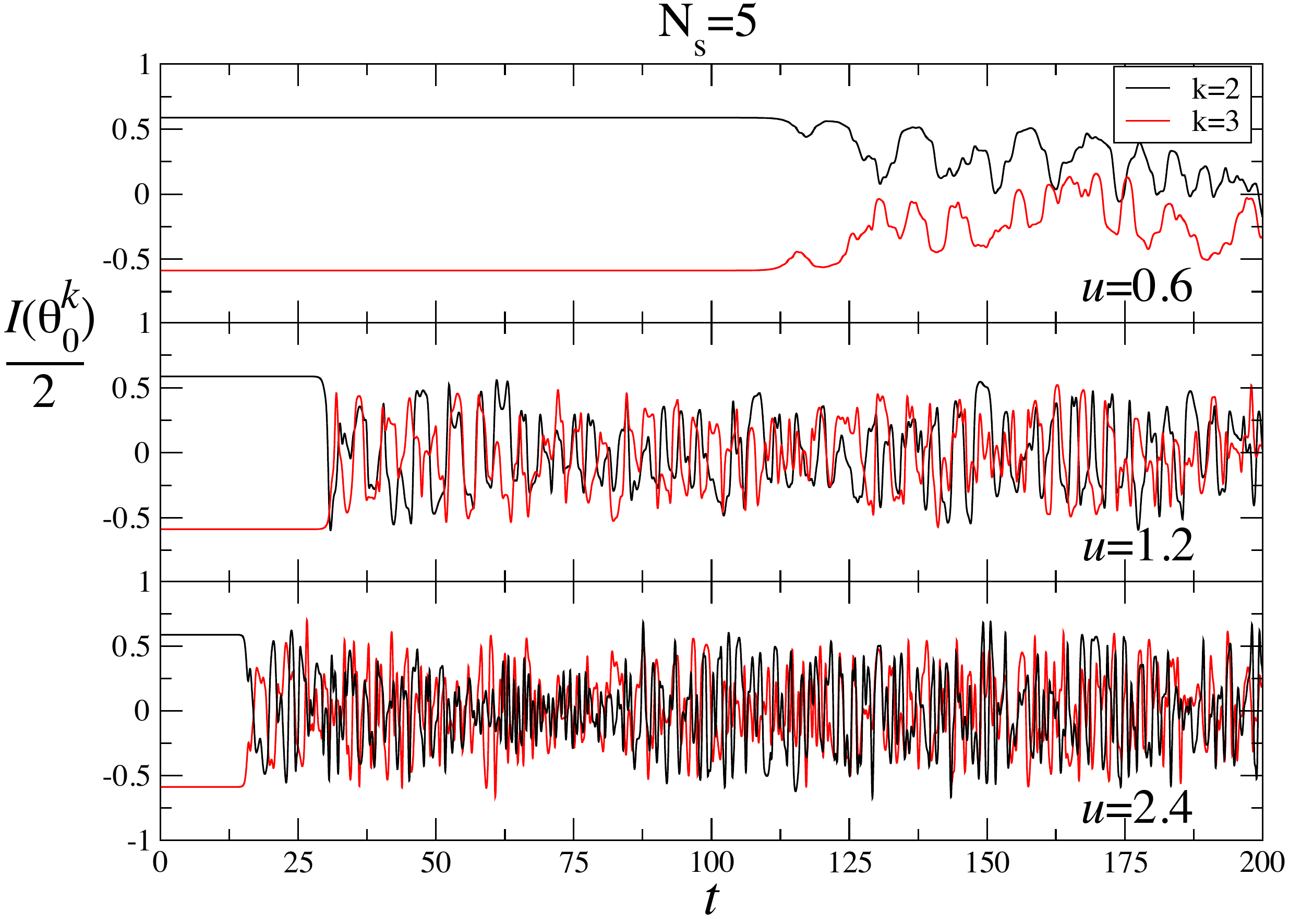} 
\end{center}
\caption{ Time-dependent circular current \eqref{curr_ns} for homogeneous initial conditions: $N_s=5$, $\theta_0=\theta_0^{k}$, $k=2$ and $k=3$. The current is calculated numerically for three different values of dimensionless interaction $u$ shown in the panels. Time $t$ is in the units of $1/K$. }
\label{curr_ns5}
\end{figure}
\begin{figure}[b!]
\begin{center}
\includegraphics[width=0.48\textwidth]{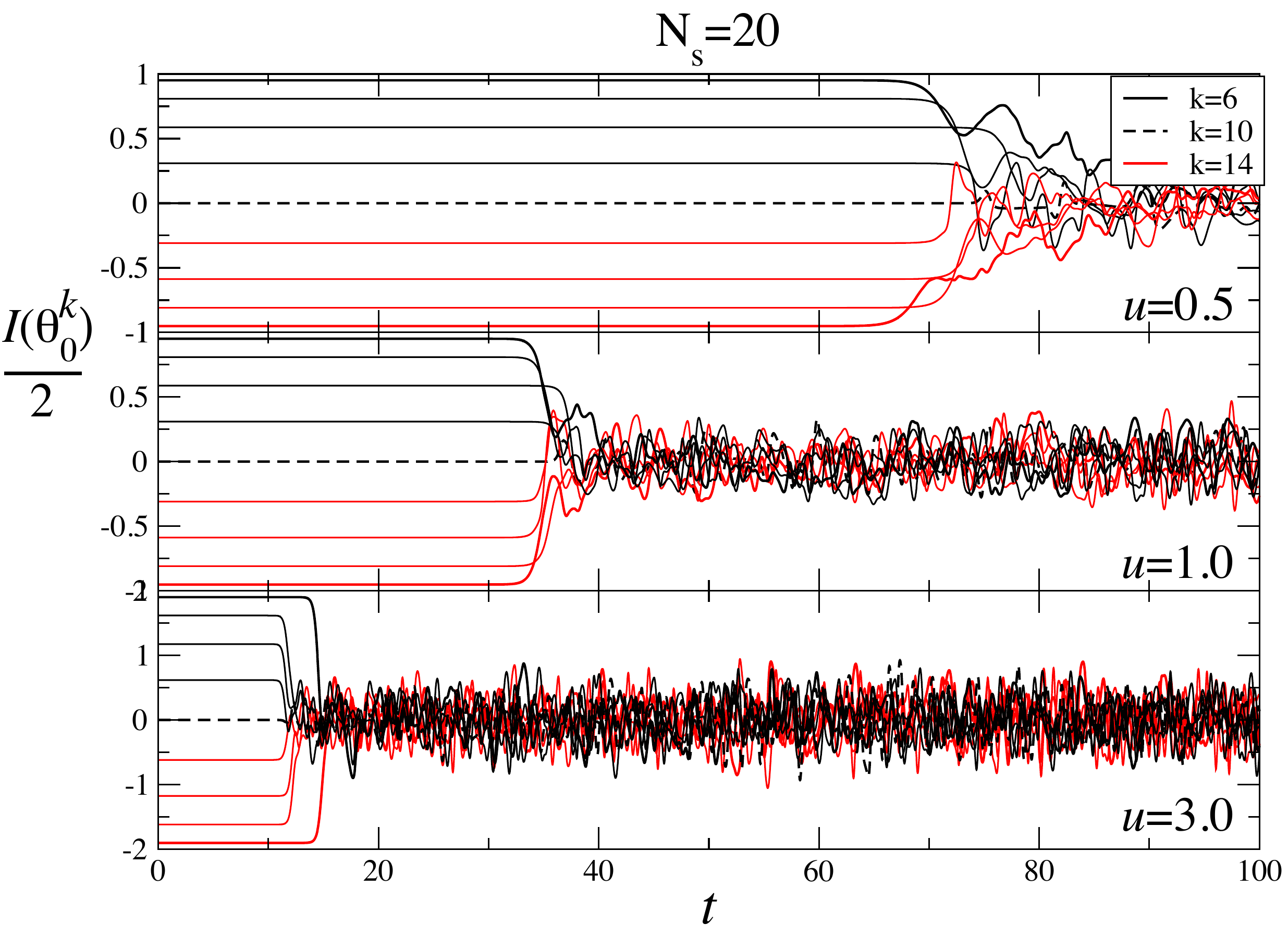}  
\end{center}
\caption{Time-dependent circular current $I(\theta_0^k)$ of the unstable modes $k=6, 7, 8, ...,14$ in the case  of $N_s=20$. 
The current is calculated numerically for three different values of $u$ shown in the panels. Time $t$ is in the units of $1/K$.  }
\label{curr_ns20}
\end{figure}

In Figs. \ref{curr_ns5}  and \ref{curr_ns20} we show examples of the time-dependent circular current of the interacting system defined in Eqs. \eqref{curr_ns} and \eqref{curr_psi}  for a ring of $N_s=5$ and $N_s=20$ condensates, respectively.  For $N_2=5$, according to Eq. \eqref{pos_modes} there are only two unstable modes $k=2$ and $k=3$. These modes are symmetric with respect to inflection at $\theta_0=\pi$ (see the inset in Fig. \ref{uc_modes}) and, therefore, the critical value of interaction $u_c$ is the same for both of the modes, $u_c\approx 0.56$. In Fig. \ref{curr_ns5} we consider three values of interaction, which are all greater than $u_c$: $u=0.6$, $u=1.2$ and $u=2.4$, and we, therefore, expect the current to behave chaotically in all three cases, which is indeed observed. However, chaos sets in at a certain time $t_c>0$ which strongly depends on the interaction $u>u_c$. One can see that  $t_c$  decreases as the interaction increases. For example, for $u$ close to $u_c$ (upper panel), $t_c$ is about 112, for $u=2.4$ it is about 15.

\begin{table}[t!]
\centering
 \caption{Values of characteristic interaction $u_c$ for different modes $\theta_0^k$ for the case of $N_s=20$. }
\begin{tabular}{ |c|c|c|c|c|c| } 
 \hline
 $k$ & 6; 14 & 7; 13 & 8; 12 & 9; 11 & 10 \\ 
 \hline
$u_c$ & 0.015 & 0.029 & 0.040 & 0.047 & 0.049  \\ 
  \hline
\end{tabular}
  \label{uc_table}
\end{table}

To find out whether $t_c$ has a mode dependence, we consider in Fig.~\ref{curr_ns20} a larger ring of $N_s=20$. It has 9 discrete, unstable modes whose critica linteraction strengths $u_c$ are listed in table \ref{uc_table}. The circular currents are shown for different values of $u$, all greater than the respective $u_c$. 
We observe that $t_c$ does not only depend on $u$, but also on the mode. For example, for $u=3$ values of $t_c$ of the "outer" modes ($k=6$ and $k=14$) are clearly greater than $t_c$ for other modes. 

\begin{figure}[!bt]
\begin{center}
\includegraphics[width=0.46\textwidth]{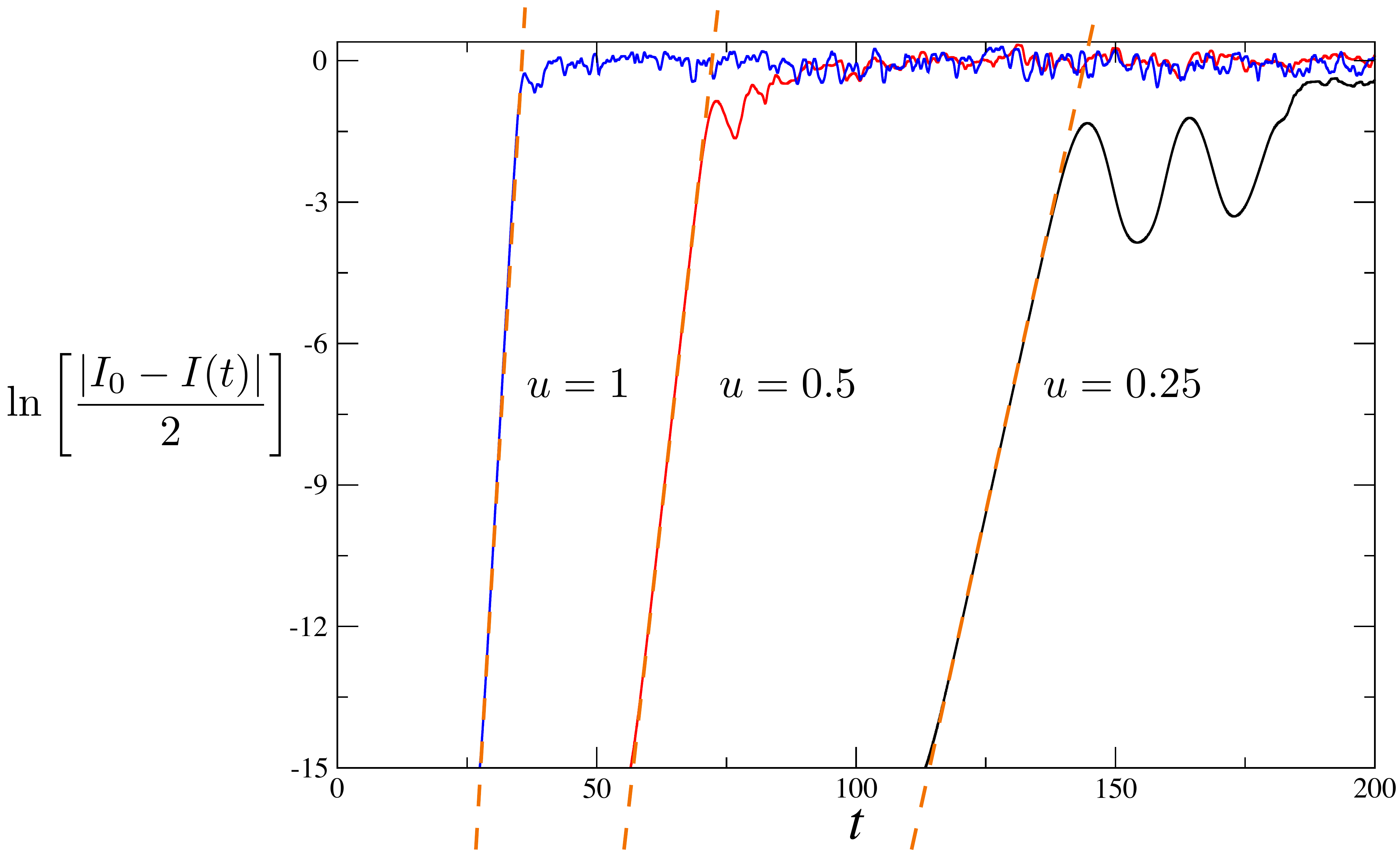}  
\end{center}
\caption{ Logarithmic plot of the deviation of the current from its initial value $I_0$ versus time for $N_s=20$, $k=6$ and three different $u$-s.  Time $t$ is in the units of $1/K$. Dashed lines are linear fits of linear parts of the graphs with intercept $b$ and slope $\alpha$: $\alpha\approx 0.5$, $b\approx -71.62$ (for $u=0.25$); $\alpha\approx 0.98$, $b\approx -70.55$ ($u=0.5$); $\alpha\approx 1.98$, $b\approx -71.62$ ($u=1$).   }
\label{log}
\end{figure}

In order to quantify the onset of chaos, we show in Fig.~\ref{log} the time evolution of the deviation of the circular current $I(t)$ from its noninteracting value $I_0(\theta_0^k)\equiv I_0$ on a logarithmic scale. As expected from the linear stability analysis, this deviation is initially exponential in time and, thus, initially the current deviation remains exponentially small, too small to be resolved in the linear plots of Figs.~\ref{curr_ns5} and \ref{curr_ns20}. 

The exponential behaviour of the deviation is governed by two parameters $\alpha$ and $b$
\begin{equation}
   \Delta I(t) \equiv |I_0-I(t)|/2 =  \ e^{\alpha t+b}\, , 
    \label{current_exp}
\end{equation}
which can be determined from 
linear fits to the logarithmic plot within the exponential time range (see dashed lines in Fig.~\ref{log}). The comprehensive analysis for a wide range of 
system paramters $u\in [0.5, 10]$ and initial conditions $\theta_0^k\in [0,\pi]$ shows that 
the coefficient $b$ has no systematic dependence on system parameters, 
with an average value of $\langle b\rangle \approx -72.34$ and a standard deviation of $\Delta b \approx 1.51$.  
Furthermore, from Fig.~\ref{log} and similar plots for the wide parameter range mentioned above (not shown), we find that chaotic behavior, 
i.e. deviation from the exponential time evolution, sets in abruptly when the  normalized current deviation $\Delta I(t)$ reaches a universal value of about 
$\ln [\Delta I(t_c) ] \equiv \lambda=-2$. This defines, via Eq.~\eqref{current_exp}, the critical time for onset of chaos as,
\begin{equation}
    t_c\approx \frac{\lambda-b}{\alpha}\approx \frac{70}{\alpha}. 
    \label{tc_estimate}
\end{equation}
It shows that both, the initial exponential deviation and the onset of
chaotic evolution, are controlled by the instability exponent $\alpha$
alone, where, however, the critical time $t_c$ is a factor of 70 larger than the exponential time scale $1/\alpha$. We note in passing that similar long-time coherent evolution with abrupt chaos onset was found in single Bose-Josephson junctions \cite{Trujillo2009}. Thus, the observed dependence 
of $t_c$ (see Figs.~\ref{curr_ns5}, \ref{curr_ns20}) on system parameters
enters through the dependence of $\alpha$, which we explore next.

\begin{figure}[!bt]
\begin{center}
\includegraphics[width=0.5\textwidth]{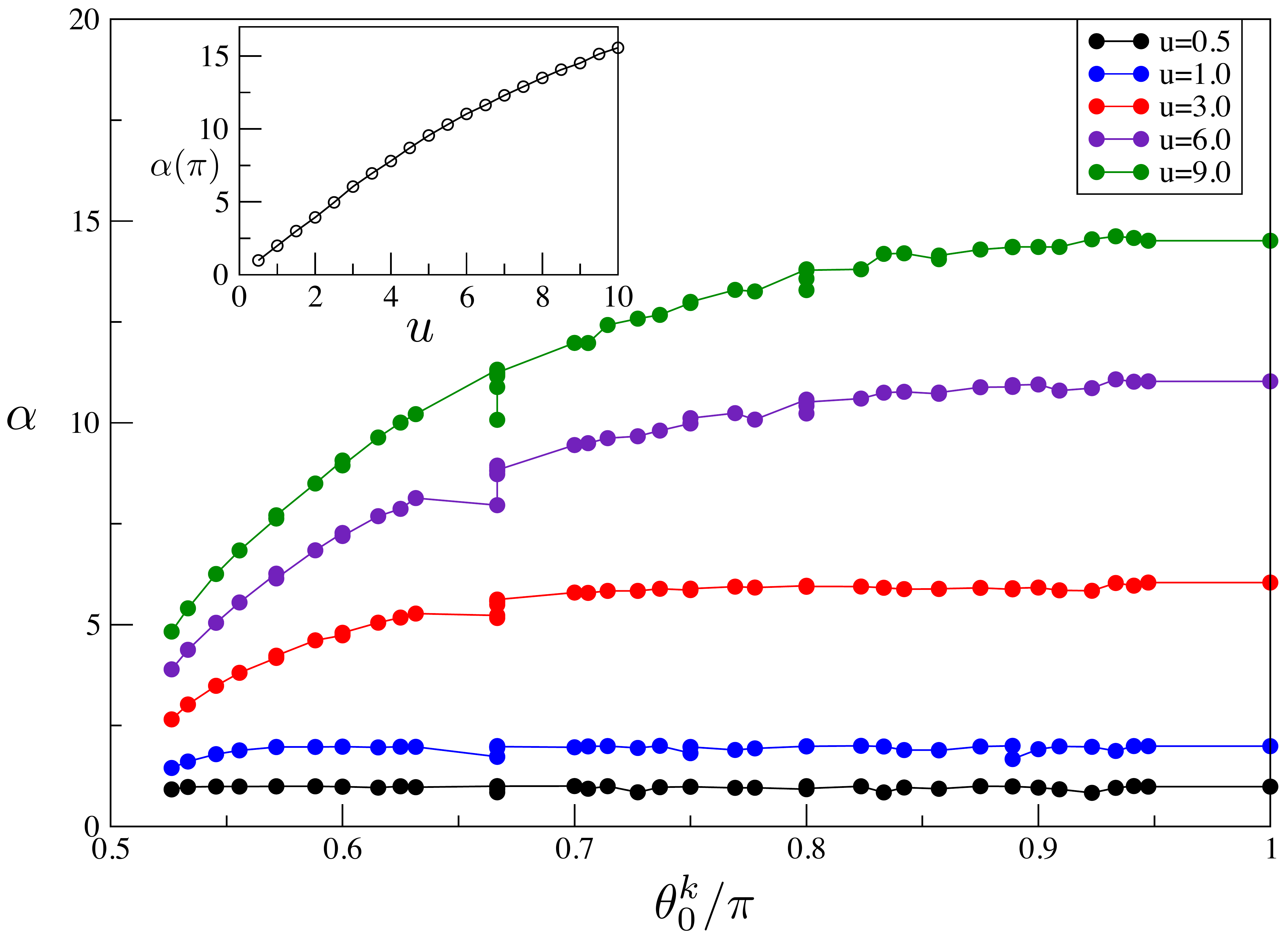}  
\end{center}
\caption{ The exponent $\alpha$ versus initial $\theta_0^k$ for different values of $u$. The inset shows the dependence of the maximum value of $\alpha$  equal to $\alpha(\theta_0^k=\pi)$ versus $u$.   }
\label{alpha}
\end{figure}

In Fig. \ref{alpha} we show results of the numerical evaluation of $\alpha$ depending on initial mode $\theta_0^k$ and interaction $u$. The plot contains data for various $N_s$, $k$ and interaction $u$. Short vertical lines of data at $\theta_0^k=2\pi/3\approx 0.666\, \pi$ correspond to rings with  $N_s=3,6,9...$, because they all have such a mode. However, the dependence of a particular mode on $N_s$  is very weak, so that we neglect it in the discussion and analysis. As expected, $\alpha$ tends to zero for $\theta_0^k\rightarrow \pi/2$, since this value of  $\theta_0^k$ marks the stability boundary (see the stability diagram in Fig. \ref{uc_modes}). Away from the stability boundary, i.e. close to $\pi$-modes, $\alpha$ has a weaker dependence on $\theta_0^k$ and is significantly dependent on $u$. This dependence is approximately linear, as shown in the inset: $\alpha(\pi)\approx 2 u$ on the interval $u\in (u_c,5)$, with $u_c\ll 1$ in this case. For small $u$, $\alpha$ tends to zero, i.e., the modes become stable.

At this point, it is interesting to compare the numerically estimated $\alpha$ with the analytical value from \eqref{alpha_analytical}. It turns out that the agreement holds only for small rings  ($N_s=3$, $N_s=4$), whereas for larger rings the agreement with this relation quickly degrades and holds only in the close vicinity of $u_c$. This defines the applicability of linear stability analysis in our systems - essentially only in the vicinity of $u_c$. 

\begin{figure}[!bt]
\begin{center}
\includegraphics[width=0.5\textwidth]{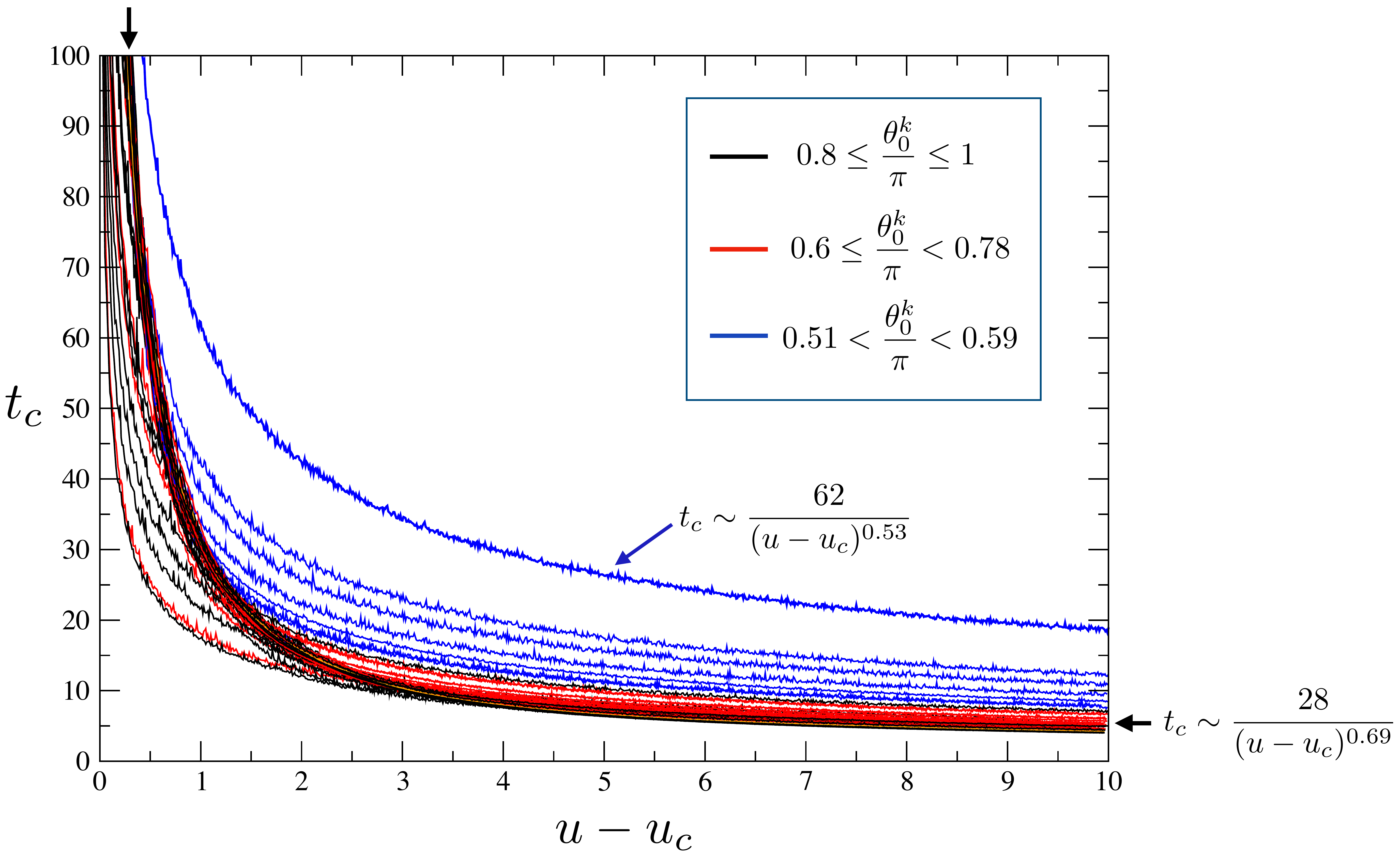} 
\end{center}
\caption{ Dependence of the onset of chaotic behavior  $t_c$ on interaction $u$ measured from $u_c$ for all unstable discrete modes in the interval $(\pi/2,\pi]$ for $3\le N_s\le 20$. The dashed curve is for $N_s=47$ and $k=12$ (this amounts to $\theta_0\approx 0.511 \pi$).  Initial conditions are specified by $\theta_0^k$. We split the initial conditions into color-coded intervals.    }
\label{all}
\end{figure}

\begin{figure}[b!]
\begin{center}
\includegraphics[width=0.43\textwidth]{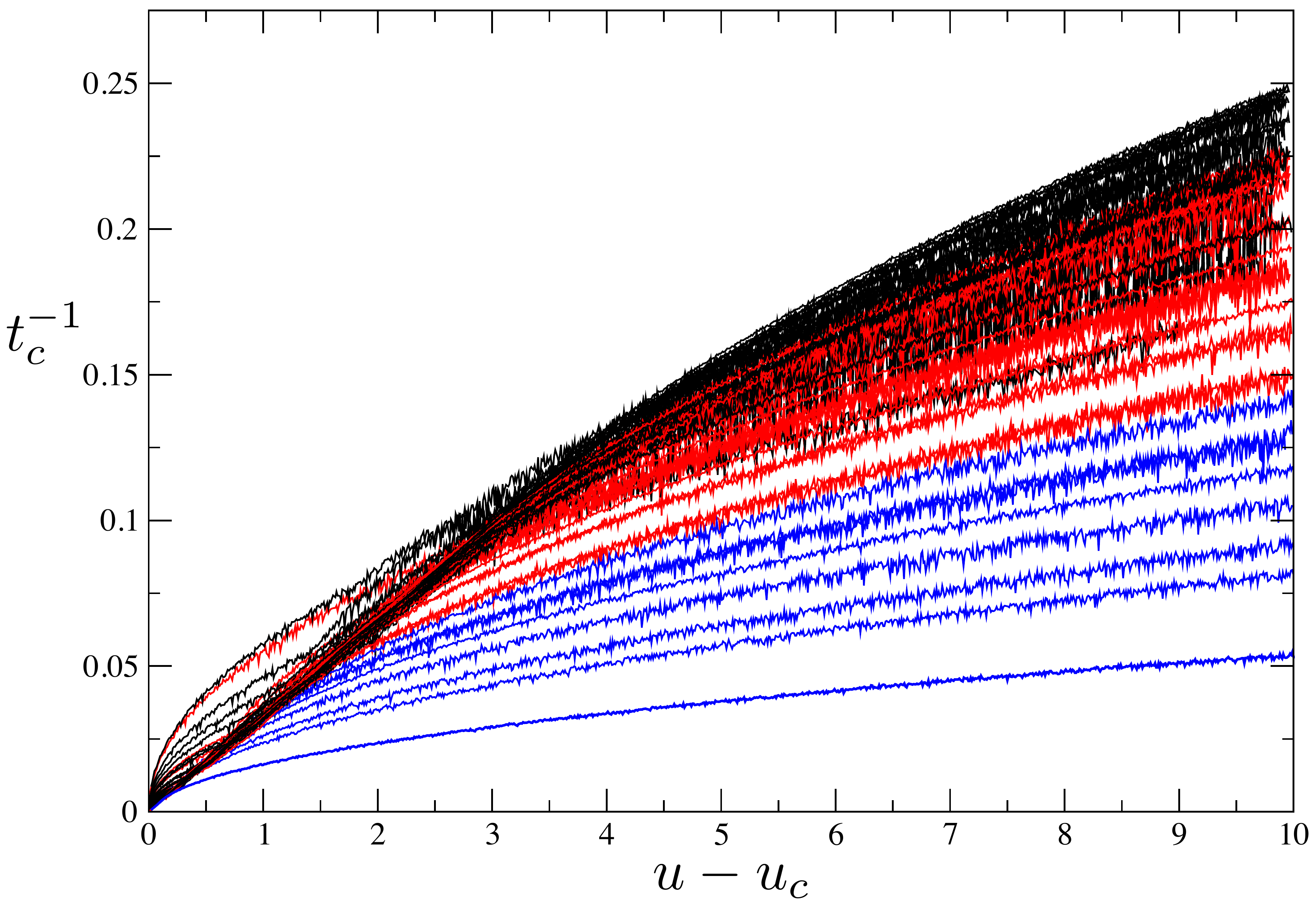} 
\end{center}
\caption{ Dependence of the inverse onset of chaotic behavior  $t_c^{-1}$ on interaction $u-u_c$ for all unstable modes in the interval $(\pi/2,\pi]$ for $3\le N_s\le 20$. The lowest curve is for $N_s=47$ and $k=12$. The curves are color-coded according to initial conditions in the same way as in Fig. \ref{all}.   }
\label{inverse_tc}
\end{figure}

Coming back to the chaos onset time $t_c$, as expected, it becomes arbitrarily large in two cases: close to stability boundaries in terms of initial phase difference, i.e. close to $\theta_0=\pi/2$ and $\theta_0=3\pi/2$, and for interactions close to $u_c$. This can be seen in Fig. \ref{all}, where we present numerically calculated  $t_c$ as a function of $u$ for a broad range of initial conditions $\theta_0^k$,   in particular, for all unstable modes $\theta_0^k\in(\pi/2,\pi]$ of all rings between $N_s=3$ and $N_s=20$.  To not overload the graph with information, we color-coded curves according to their initial conditions split into intervals of $\theta_0$: "black" corresponds to $\theta_0^k\in[0.8\pi,\pi]$, closest to the anti-phase mode and farthest from the stable modes.  One can see that most of them (in particular those with small $u_c$) bunch around the curve marked by the black arrows. The marked curve is easily fit with a two-parameter fitting function resulting in  $t_c\approx 28/(u-u_c)^{0.69}$. 
"Red" curves represent the next interval $\theta_0^k\in [0.6 \pi, 0.78 \pi)$. They start to deviate from the black curves, especially for large $u$. This can be better seen in Fig. \ref{inverse_tc}, where we plot $t_c^{-1}$ versus $u-u_c$. 
Last, the blue curves span the interval of $\theta_0^k\in [0, 0.59 \pi]$. 
One curve is calculated for $N_s=47$ and $k=12$, being an example of a curve relatively close to $\pi/2$-mode with $\theta_0^{k}\approx 0.511$ (this is the uppermost curve in the graph). The fitting of this curve results in $t_c\approx 62/(u-u_c)^{0.53}$, leading to a sizable $t_c$ even for large values of $u$.

Note that $t_c$ can be determined differently, for example, from the deviation from its initial value by $0.0001$ of any site occupation. Finding $t_c$ from site occupation would lower the $t_c$ of entire Fig.\ref{all} about an average of $10\%$, which does not change any conclusions or analysis. 

To get an idea of how large $t_c$ can be in terms of experimental values, we translate the value $t_c=100$ to ms from experimental data on arrays of condensates \cite{Cataliotti2001}. For example, from experiments on long arrays of cold atoms ($N_s=200$), we take the value of Josephson coupling $K\sim 0.07 E_R$, where $E_R$ is the recoil energy of $^{87}$Rb atom of mass $m$ absorbing one of the lattice photons
\begin{equation}
    E_R=\frac{h^2}{2m\lambda^2}\sim 1.5\cdot 10^{-11} eV
\end{equation}
with $\lambda=795\cdot 10^{-9}$m, \ \cite{Cataliotti2001}. This gives us the value of $K\sim 1.05 \cdot 10^{-12}$eV. From this analysis it follows that the dimensionless $t_c=100$ will correspond to $62.86$ ms. Such values of $t_c$ could be achievable for very weak interactions as follows from Fig. \ref{all}. 

\begin{figure}[t!]
\begin{center}
\includegraphics[width=0.4\textwidth]{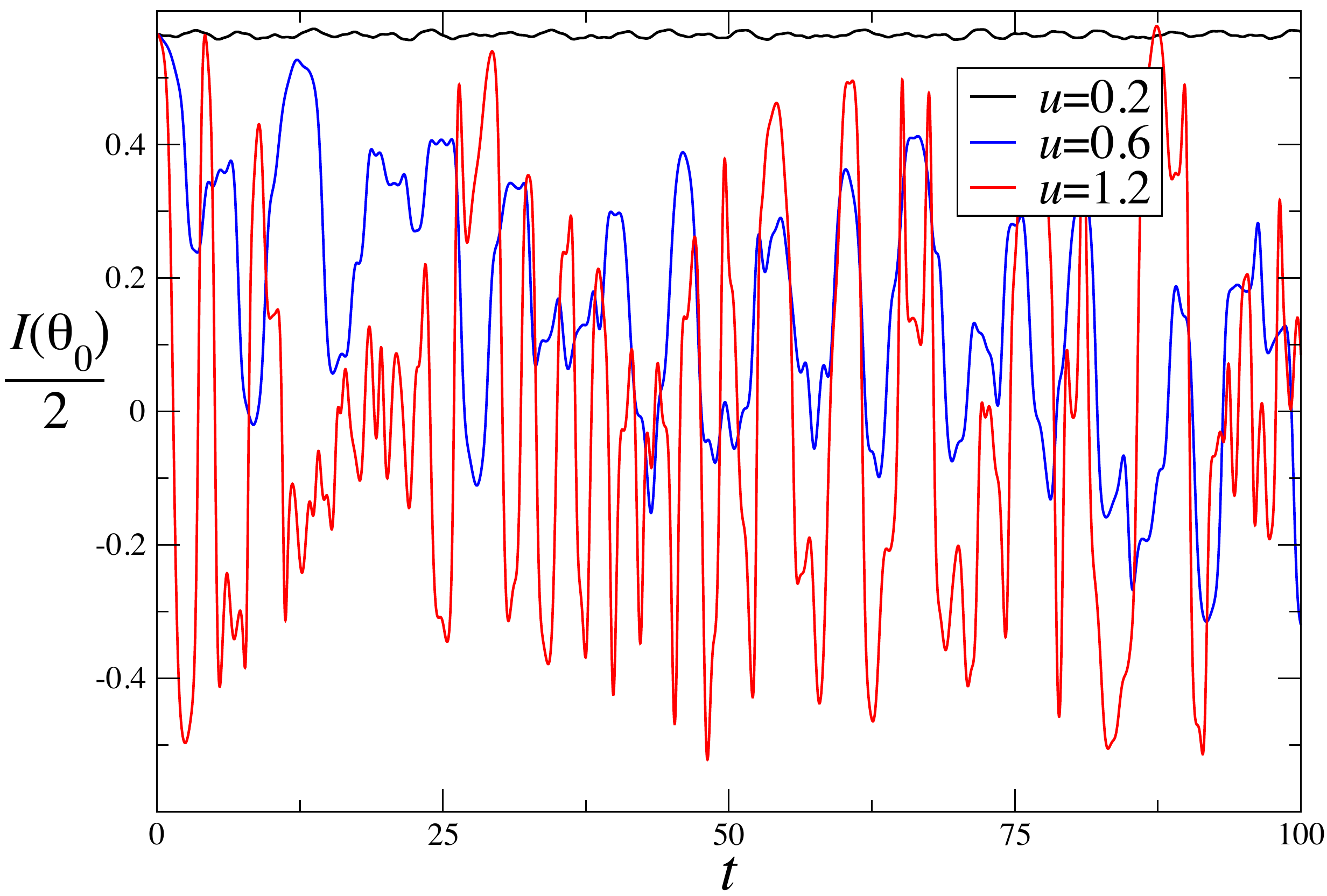} 
\end{center}
\caption{Current oscillations with time for initial conditions  $\theta_0=0.75 \pi$ and $N_s=20$, i.e. for a non-eigenmode. The results are shown for three different values of dimensionless interaction $u$ shown in the inset.}
\label{custom_theta}
\end{figure}

Note, that when initial conditions fall in between the eigenmodes, one expects time-dependent current, as shown in Fig. \ref{custom_theta} for $\theta_0=0.75 \pi$ and $N_s=20$. We see that with increasing $u$ and therefore non-linearity, current oscillations become larger, more chaotic and eventually average to zero over time for large $u$, as we will see in the next section. The notion of $t_c$ does not make sense in this case, since there appears no additional energy scale associated with real parts of eigenvalues. 

\subsection{Time-averaged circular current and coherence}

\begin{figure}[t!]
\begin{center}
\includegraphics[width=0.48\textwidth]{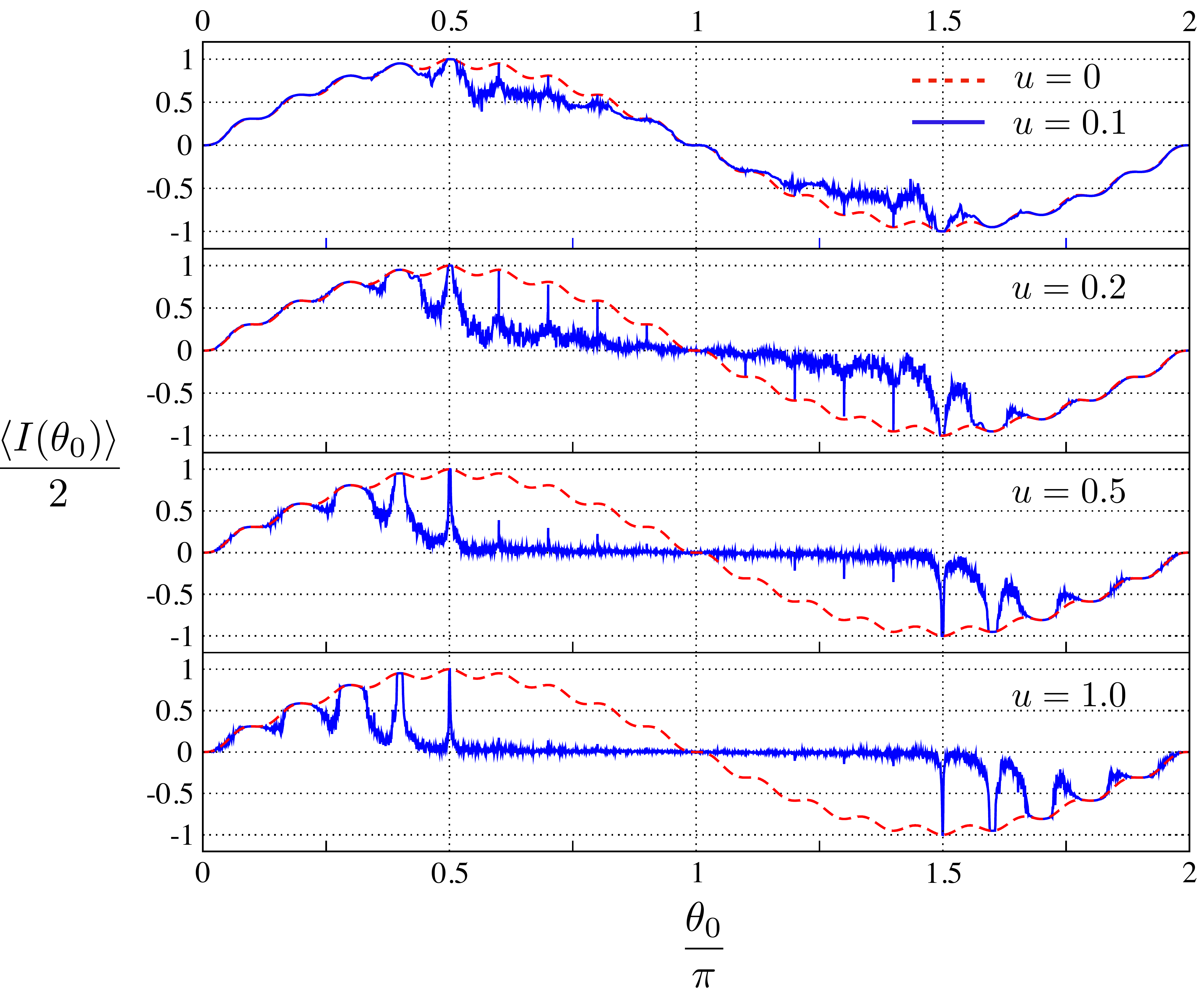} 
\end{center}
\caption{ Time-averaged circular current normalized by its maximum value (=2) versus initial phase difference $\theta_0$. The current is calculated for $N_s=20$ ring and four different values of $u$ indicated in the panels. The red dashed line corresponds to the $u=0$ non-interacting case and is put as a reference, the solid lines correspond to the interacting systems. The current is averaged over the time interval $t\in [0,200]$.}
\label{average_curr}
\end{figure}

We now evaluate the time-averaged circular current numerically 
\begin{equation}
\langle I (\theta_0)\rangle =\lim_{T\rightarrow \infty}\frac{1}{T}\int dt\,  I(\theta_0)
\label{curr_av}
\end{equation}
for initial conditions \eqref{initial} with $\theta_0$ ranging from $0$ to $2\pi$ covering stable and unstable regions of the stability diagram in full. We chose a ring with $N_s=20$ sites and four values of $u$ greater than the maximum $u_c$ listed in Table I. Given the discussion about $t_c$ in the previous section, it is clear that the resulting time-averaged current at the discrete unstable modes will depend on the numerically available time interval, over which we can let our program run, providing reliable results.  In our case $t\in [0,200]$. The results for the averaged current are presented in Fig. \ref{average_curr}. 
 The four panels of Fig. \ref{average_curr} correspond to four different values of $u$ as indicated. The dashed curve is the current for $u=0$ and is shown on all the panels for reference. For small values of the dimensionless interaction $u=0.1$ the deviations of the current from $u=0$ values in the stable part of the plot are hardly visible (i.e. for $\theta_0\in [0,\pi/2]$ and $[3\pi/2,2\pi]$). For the unstable part, $\theta_0\in(\pi/2,3\pi/2)$, it is striking that $t_c$ is, in fact, greater than $200$, otherwise the averaged current at the unstable discrete modes stays unaffected by the chaotic regime even though $u>(u_c)_{max}$. Since there are no additional time scales for $\theta_0$ in between the discrete modes, the average current values begin to deviate from their $u=0$ values due to chaotic dynamics (see Fig. \ref{custom_theta}). 

Upon increasing interaction, $t_c$ decreases, so that for $u=0.2$ the $t_c$ becomes of the order of $180$, which is comparable with the maximum value of $t$ in our numerics. As a result, the current values at discrete unstable modes remain at most unaffected, whereas the in-between $\theta_0$-s correspond to circular currents quickly averaging out to zero. This effect is also visible around the stability boundaries $\theta_0=\pi/2$ and $\theta_0=3\pi/2$. Although current values at stable discrete modes remain constant, the values in between begin to feel the effect of the increasing interaction. These tendencies become more pronounced as we increase the interaction further. As $t_c$ rapidly decreases ($t_c\sim 75$ for $u=0.5$ and $t_c\sim 35$ for $u=1$), so does the time-averaged current in the unstable region. For $u=1$, this current is practically zero everywhere in the unstable region.  The stable region is less affected by interaction, and the effect is most visible for values of $\theta_0$ falling in between the stable discrete modes, close to the stability boundaries. In the limit of large $N_s$, the unstable modes fill the interval $(\pi/2,3\pi/2)$, while the stable modes fill the intervals $[0,\pi/2]$ and $[3\pi/2,2\pi]$. Large $N_s$ approaches the continuous limit where all $\theta_0$ are eigenmodes (because $\theta_0^k$ merge together). The current in the limit of large $N_s$ is described by $2 \sin\theta_0$, but in the unstable region the current averages to zero above  $u_c$, provided averaging time period is greater than $t_c$.

\begin{figure}[!bt]
\begin{center}
\includegraphics[width=0.48\textwidth]{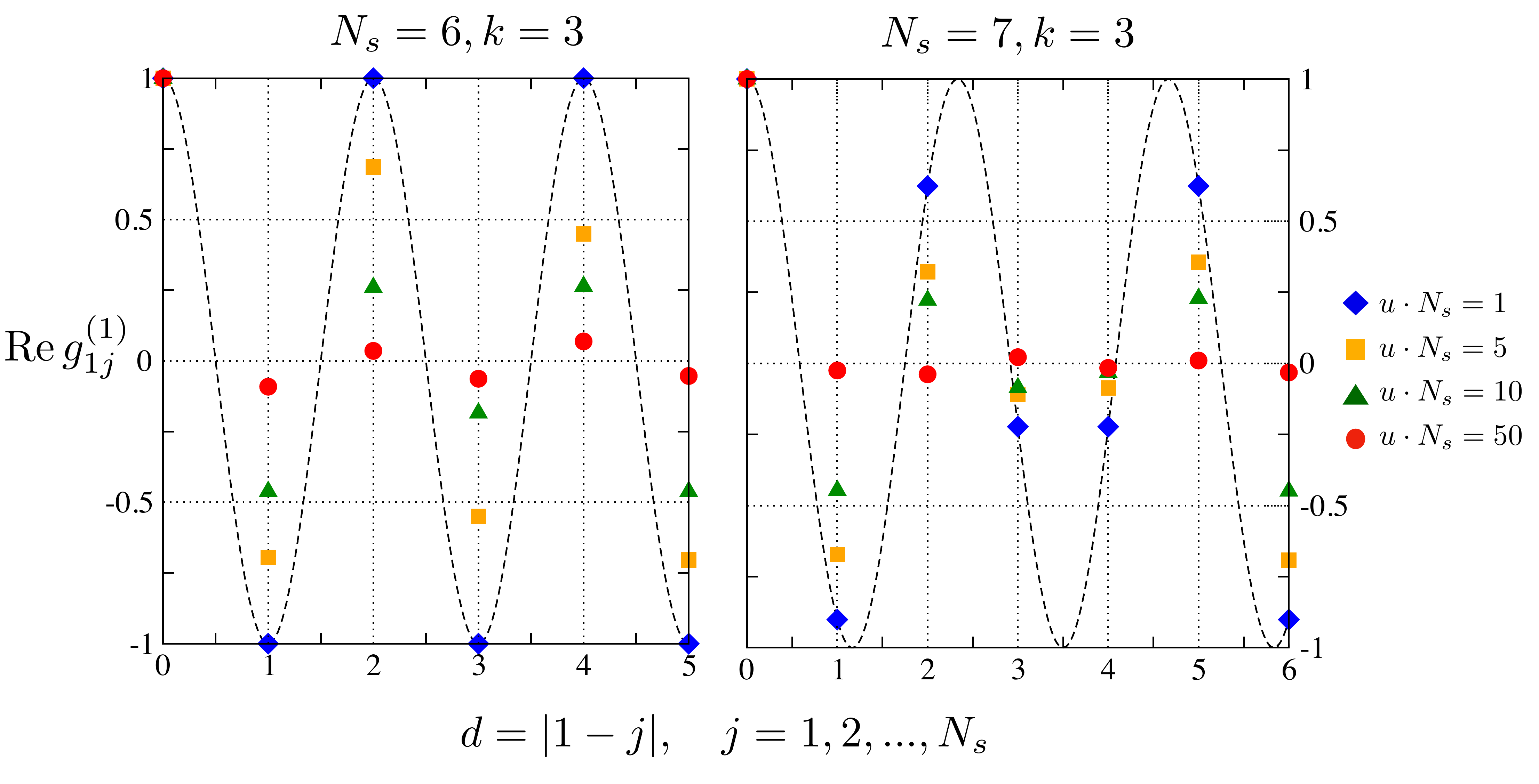} 
\end{center}
\caption{ The real part of the first-order correlation function Re$\, g_{ij}^{(1)} $ averaged over time on the interval $t\in[0,200]$. The correlation function is plotted as a function of distance w.r.t. to the first site (the distance is normalized by a lattice constant $a\equiv 1$).     }
\label{coherence}
\end{figure} 

The circular current averaging to zero is equivalent to a loss of coherence in the system. The coherence can also be quantified by the first-order coherence function 
\begin{equation}
    g_{ij}^{(1)}=\frac{\langle \psi_i^* \psi_j\rangle }{\sqrt{\langle |\psi_i|^2 \rangle \langle |\psi_i|^2 \rangle}}.
\end{equation}
It can be experimentally measured by interferometry as a function of a distance from a fixed site. In Fig. \ref{coherence} we show the real part of the coherence function versus distance from site $i=1$.  We chose two different initial values of the phase difference: $\theta_0^{(3)}$ for $N_s=6$, i.e. the $\pi$-mode of a 6-site ring, and a mode $\theta_0^{(3)}$ close to $\pi$ of a 7-site ring. Although both of these modes are unstable, experiments on polariton condensates \cite{Cookson2021} demonstrate nonzero circulating currents and coherence corresponding to small values of $u$, e.g. $u N_s=1$ in Fig. \ref{coherence}. It means that either the interaction was rather small in the experiment, and $t_c$ relatively large, or there were other stabilizing factors in the experimental system. For example, the experimental system is an open system, whereas our system is closed. 

In Fig. \ref{coherence} we further demonstrate how the coherence is destroyed by interaction and averages to zero for increasing $u$ (see data for $u\cdot N_s=10$ and $u\cdot N_s=50$).

\section{Conclusions and discussion}

\label{last}

We analyzed circular currents and their stability in rings of condensates under specific initial conditions of equal filling and homogeneous phase differences. We found a set of discrete eigenmodes (phase differences) differentiated by their winding numbers. When such a mode falls into the interval $(\pi/2,3\pi/2)$, it is stable until the interaction exceeds a certain value $u_c$. This critical interaction depends on the mode and on the ring size. When $u_c$ is exceeded, the system dynamics and the circular current become chaotic, with the current quickly averaging to zero over time. This marks the effective loss of coherence in the system. We showed that this dephasing occurs not immediately upon entering the unstable regime but rather after a chaos onset period $t_c$. This period $t_c$ can be arbitrary large when the system is close to the stability boundary, i.e. when the unstable mode under consideration is close to $\pi/2$ or $3\pi/2$, or when $u$ is close to $u_c$. For modes close to the antiphase $\pi$ mode in the instability region, i.e. for modes falling onto interval $[\pi,\pi\pm0.2\pi]$ one can even establish a universal behavior of $t_c\approx 28 (u-u_c)^{-0.69}$.

We also established that the critical time for chaos onset, that is, the dephasing time, is proportional to but about two orders of magnitude larger than the time scale of exponential deviations, $1/\alpha$, where $\alpha$ is the current instability exponent. This may be relevant for technological applications of quantum coherent dynamics of rings of interacting Bose-Einstein condensates. The presence of the large chaos onset period $t_c$ explains why in the previous works about three-site condensate rings, the circular current was non-vanishing in the chaotic regime. It may also shed light on recent experimental observations of circulating currents in loops of polaritonic condensates with large winding numbers, although there can be other stabilizing factors. 

In the future, it would be interesting to explore how fluctuations would affect the present description and extension of the work to polaritonic condensates would be of interest.  

\vspace{1cm}

\section*{Acknowledgments}

We acknowledge S. Ray and especially M. Eschrig for many fruitful discussions regarding the project. We are also grateful to an anonymous Referee for their excellent comments which helped us improve the manuscript.  This work was financially supported in part (J.K.) by the Deutsche Forschungsgenmeinschaft (DFG) via SFB/TR 185 (277625399) and the Cluster of Excellence ML4Q (EXC $2004/1-390534769$).

\section*{Appendix A: Exact solution of the non-interacting case.  }

The non interacting case when $u=0$ in Eq.\eqref{GPsystem} can be solved analytically by rewriting these simultaneous equations in a matrix form. Albeit it cannot be written in a succinct final solution for $\psi$ due to a large number of unknown coefficients. However with the simplifications of $\hbar\equiv 1$, $E_i\equiv E_0$, $U_i\equiv U$, $K_{i,i+1}\equiv K$ for  $i=1,..,N_s$ and expressing the time argument $t$ in the units of $1/K$ we get 
\begin{eqnarray}
\mathrm{i} \frac{\partial}{\partial t}\Vec{\psi}(t)= \hat H \Vec{\psi}(t)
\end{eqnarray}
where $\epsilon = E_0/K$, vector $\Vec{\psi}=(\psi_0,\psi_1, .., \psi_{N_s-1})$ contains $N_s$ condensate functions which are numbered from $0$ to $N_s-1$ for convenience,  and  $\hat H$ is a $N_s\times N_s$ matrix
\begin{equation}
    \hat H=\begin{pmatrix}
     \epsilon & -1 & 0 & 0 & \dots & 0 & -1 \\
     -1 & \epsilon & -1 & 0 & \dots & 0 & 0 \\
     0 & -1 & \epsilon & -1 & \dots & 0 & 0 \\
     & & & \ddots & & & \\
      0 &  0 & 0  & 0  & \dots & \epsilon  & -1 \\
     -1 & 0 & 0 & 0 & \dots & -1 & \epsilon 
    \end{pmatrix}.
    \label{ham_circ}
\end{equation}
This matrix is circulant and real-symmetric with real eigenvalues (in our case, coinciding with energy eigenvalues). The eigenvectors $\Vec{v}$ of circulant matrices are well-known and do not depend on circulant matrix entries \cite{Circulant_matrix}. The $k$-component of $\vec{v}$, corresponding to an eigenvalue $\lambda_j$ reads
\begin{equation}
\label{circulant_e_vec}
 v_{k}(\lambda_j) = \exp{\mathrm{i} \frac{2 \pi j k } {N_s}}.
\end{equation}
Here indices $k, j =0,1,...,N_s-1$. 
The eigenvalues of $\hat H$ are also readily found, as they are the discrete Fourier transforms of the first row of the matrix $\hat H$
\begin{equation}
\label{circulant_e_val}
\lambda_j =  \sum^{N_s-1}_{k=0} h_{0k} \exp{\mathrm{i} \frac{2 \pi j k } {N_s}}, 
\end{equation}
where $h_{00}, h_{01}, ...,  h_{0,N_s-1} $ are the entries of the first row of the circulant matrix \eqref{ham_circ}. In our case the eigenvalues acquire a simple form 
\begin{equation}
 \lambda_j= \epsilon - 2\cos{\left( \frac{2 \pi j}{N_s } \right)}. 
\end{equation}

This gives us a general solution
\begin{equation}
    \psi_{n}(t)=\sum_{j=0}^{N_s-1}c_j v_{n}(\lambda_j) \, e^{-\mathrm{i} \lambda_j t},
\end{equation}
where the coefficients $c_j$ are defined by initial values of the condensate wave-functions $\vec{\psi}(0)$ 
\begin{equation}
 \vec{\psi}(0)=\hat F \vec{c}
\end{equation}
where the discrete Fourier transform matrix $\hat F$ contains eigenvectors $\vec{v}(\lambda_j)$ as columns
\begin{equation}
    F_{kj}=v_k(\lambda_j)
\end{equation}
with $k,j=0,1,...,N_s-1$. $\vec{c}$ is hence determined by the inverse Fourier transform

\begin{equation}
    c_j=\frac{1}{N_s}\sum_{k=0}^{N_s-1}e^{-\mathrm{i} \frac{2\pi }{N_s}k\cdot j} \psi_k(0). 
\end{equation}

Finally,

\begin{eqnarray}
\psi_n (t) = \frac{1}{N_s}\sum_{j,k=0}^{N_s-1}   e^{ \mathrm{i} \frac{2 \pi }{N_s} (n-k)j }   \psi_k(0) e^{-\mathrm{i} \lambda_j t}  . 
\end{eqnarray}

For the particle number per site $N_n = \psi_n \psi_n^*$ we get
\begin{eqnarray}
N_n &=& \sum_{j,k=0}^{N_s-1 } c_{n,j,k} \ e^{-\mathrm{i} t ( \lambda_j - \lambda_k)}, \\
c_{n,j,k} &=& \frac{1}{N_{s}^{2}}\sum_{l,m=0}^{N_{s}-1 }  e^{ \mathrm{i} \frac{2 \pi }{N_{s}} (n-k-l+m)j }   \psi_{l}(0) \psi^*_{m}(0).
\end{eqnarray}

As an example, we present solutions for $N_s=3$. The eigenvalues are $\lambda_1=\lambda_2 = \epsilon + 1$, $\lambda_3 = \epsilon-2$ and the three wavefunction have the form
\begin{equation}
\label{psi_3}
\psi_i(t) = \frac{S}{3} e^{ \mathrm{i} (2-\epsilon)  t} + \left(\psi_i(0)-\frac{S}{3}\right) e^{- \mathrm{i} (\epsilon + 1) t},\quad S = \sum_{i=1}^{3} \psi_i(0).
\end{equation}
The occupation number per site is then 
\begin{equation}
\label{ni_3}
n_i = n_i(0) + 2\Re \left[ \frac{S}{3}\left(\psi^*_i(0)-\frac{S^*}{3}\right) \left( e^{3 \mathrm{i} t} - 1 \right)\right].
\end{equation}
For initial conditions \eqref{initial} and $\theta_0$ coinciding with the eigenmodes \eqref{loc_max}, the $S=0$ giving the simple expressions
\begin{eqnarray}
 \psi_i(t) & =& \psi_i(0) e^{- \mathrm{i} (\epsilon + 1)  t}, \nonumber \\
 n_i(t)& =& n_i(0)= 1.
\end{eqnarray}

\section*{Appendix B: Linear stability analysis of the interacting system.  }

Eqs. \eqref{eqs_fin} are real equations describing the dynamics of interacting  system. In order to analyze their linear stability we construct the Jacobian matrix for $2N_s$ variables $n_1,n_2,...,n_{N_s}, \varphi_1,\varphi_2,...,\varphi_{N_s}$
\begin{equation}
J = 
\begin{pmatrix}
\dv{\dot{n}_1}{n_1}   &  \dots &  \dv{\dot{\varphi}_{N_s}}{n_1}   \\
\vdots      &  \ddots   &    \vdots     \\
\dv{\dot{n}_1}{\varphi_{N_s}}     &  \dots    &     \dv{\dot{\varphi}_{N_s}}{\varphi_{N_s}}    
\end{pmatrix}
\end{equation}
where $\varphi_i = \theta_{i,i+1} $.  

Fixed points are determined from the steady state condition of Eqs.\eqref{eqs_fin} 
\begin{equation}
x=n_{i} = 1 \ , \ \ y=\varphi_{i} = \frac{2\pi}{N_s} k , \quad i,k =  1, ..., N_s.
\end{equation}
The  Jacobian matrix $J$ at the fixed points can be written as a two by two block matrix 
\begin{equation}
J = 
\begin{pmatrix}
 S & C \\
 D & S \\
\end{pmatrix}.
\end{equation}
Here the matrices  $S$, $C$ and $D$ are circulant $N_s$x$N_s$ matrices, whose elements can be written as
\begin{eqnarray}
S_{ij} &=&  \sin(y)   (\delta_{i,j+1}  - \delta_{i,j-1} ), \nonumber \\
D_{ij} &=&   2x\cos(y)   ( -\delta_{i,j} +  \delta_{i,j+1} ) , \nonumber  \\
C_{ij} &=&  \left(u+\frac{3\cos(y)}{2x}\right) (  \delta_{i,j}  - \delta_{i,j-1} ) \nonumber  \\
&+& \frac{\cos(y)}{2x} \ (\delta_{i,j-2}  - \delta_{i,j+1}). 
\end{eqnarray}
We see that only $C$-matrix depends on interaction $u$. All circulant matrices of the same size have the same eigenvectors, thus all circular matrices of the same size can be diagonalized by $J_S=U^{-1}SU$ where $U$ matrix columns are the circulant matrix eigenvectors. Using the rules for inverses of block matrices, we perform a simple transformation to diagonalize the circulant matrices  in the Jacobian matrix 

\begin{equation}
\begin{pmatrix}
 U^{-1} & 0 \\
 0 & U^{-1} \\
\end{pmatrix}
\begin{pmatrix}
 S & C \\
 D & S \\
\end{pmatrix}
\begin{pmatrix}
 U & 0 \\
 0 & U \\
\end{pmatrix} =
\begin{pmatrix}
 J_S & J_C \\
 J_D & J_S \\ 
\end{pmatrix} 
\end{equation}

The similar transformation preserves the eigenvalues, the $J_S$ , $J_C$ , $J_D$ matrices are now diagonal matrices. As all of the new diagonal matrices commute, we can take advantage of a block matrix determinant rule to find the eigenvalues of $J$
\begin{equation}
Det 
\begin{pmatrix}
 J_{S}-\lambda I & J_C \\
 J_D & J_{S}-\lambda I \\
\end{pmatrix}
= Det ( (J_{S}-\lambda I)^2 -J_D J_C ) = 0
\end{equation}
This equation can be rewritten as 
\begin{equation}
 (\lambda_{j}^{S}-\lambda_{j})^2 - \lambda_{j}^{D} \lambda_{j}^{C} = 0, \quad j = 1, ..., N_s.
\end{equation}
where $\lambda_{j}^{S},\lambda_{j}^{D},\lambda_{j}^{C} $ are the eigenvalues of  $S,D,C$ respectively and $\lambda_{j}$ is the eigenvalue of  the Jacobian $J$. We get 
\begin{equation}
\lambda_{j} = \lambda_{j}^{S} \pm \sqrt{\lambda_{j}^{D} \lambda_{j}^{C}}, \quad j =  1, ..., N_s. 
\end{equation}

Now we collect the eigenvalues of the circulant matrices $S,D$ and $C$, which can be found following the method explained in Appendix A:
\begin{eqnarray}
 \lambda_{j}^{S} & = & 2 \mathrm{i} \sin(y) \sin{ \left( \frac{2 \pi j } {N_s} \right)}, \nonumber \\
\lambda_{j}^{D} & = & 2 x \cos{y} \left( \exp{-\mathrm{i} \frac{2 \pi j  } {N_s }} - 1 \right), \nonumber \\
 \lambda_{j}^{C} & = & \left(u+\frac{3\cos(y)}{2x}\right)  \left( 1 - \exp{\mathrm{i} \frac{2 \pi j  } {N_s }}  \right) \nonumber \\
 &+& \frac{\cos(y)}{2x} \left( \exp{\mathrm{i} \frac{4 \pi j  } {N_s }} - \exp{-\mathrm{i} \frac{2 \pi j  } {N_s }} \right).
\end{eqnarray}
As a result we get for $\lambda_j=\lambda_j(k)$
\begin{equation} 
\begin{split}
\lambda_{j}(k)  & = 2 \mathrm{i} \left\{  \sin \theta_0^{k}  \sin{ \left(\frac{2 \pi j } {N_s} \right) }  
\pm  \right.   \\
 & \left. \sin{ \left(\frac{ \pi j } {N_s} \right) } \sqrt{ 2 \cos\theta_0^{k} \left[2\cos\theta_0^{k} \sin^2\left( \frac{\pi j}{N_s} \right) + u  \right]    } \right\}, 
\end{split}
\end{equation}
The eigenvalues are purely imaginary if the expression under the square root is non-negative for all $j$. This would correspond to a neutral center and stable system. If at least one of the eigenvalues acquires a real part, this will correspond to an exponential instability in the system. We discuss this in more detail in the main text after Eq.\eqref{eq1}.

\end{document}